\documentclass{aa}
\usepackage{graphicx}
\begin{document}
   \title{Stellar evolution with rotation and magnetic fields:}

   \subtitle{II: General equations for the transport by Tayler--Spruit dynamo}

\author{Andr\'e Maeder, Georges Meynet}

     \institute{Geneva Observatory CH--1290 Sauverny, Switzerland\\
              email:  andre.maeder@obs.unige.ch\\
             email: georges.meynet@obs.unige.ch
               }

   \date{Received  / Accepted }

   \offprints{Andr\'e Maeder} 

   \abstract{We  further develop  the Tayler--Spruit 
   dynamo theory,  based on the most efficient instability for  generating 
   magnetic fields in  radiative layers 
   of differentially rotating stars. We avoid the simplifying assumptions 
   that either the $\mu$-- or the $T$--gradient  dominates, 
   but we treat the general case and we also account for
   the nonadiabatic effects, which favour the growth of the magnetic field.
  The general equation leads to the same  analytical solutions  in the
   limiting  cases considered by Spruit (\cite{Spruit02}).
   Numerical models of a 15 M$_{\odot}$ star with a magnetic field are performed. 
  The differences between the asymptotic solutions and the general solution 
  demonstrate the need to use the general solution. Stars
  with a  magnetic field rotate almost as a solid body.
  Several of their properties (size of the core,  MS lifetimes, tracks, abundances)
   are closer to those of models without rotation  than with rotation only.
   In particular, the observed  N/C or N/H excesses in OB stars are better explained
   by our previous models with rotation only than by the present models with magnetic fields 
   that predict no nitrogen excesses.
   
   We show that there is a complex feedback loop between the magnetic instability and the
   thermal instability driving meridional circulation. Equilibrium of the loop,
   with a small amount of differential rotation, can be  reached
   when the velocity $U_{\mathrm{magn}}$ of the growth of the magnetic instability 
   is of the same order as the velocity $U_{\mathrm{circ}}$ of the meridional circulation.
   This opens the possibility for further magnetic models, but at this stage we do not know
   the  relative importance of the magnetic fields due to 
   the Tayler instability in stellar interiors.

               }
   \maketitle
%

\section{Introduction}

A major concern is to know whether magnetic fields are important for stellar evolution
or not (cf.  Roxburgh \cite{Rox03}). Some initial numerical tests have 
 suggested that the magnetic fields 
created by the  Tayler--Spruit  dynamo (Spruit \cite{Spruit99},
\cite{Spruit02})
in rotating stars may be large enough  to transport 
  angular momentum in stellar interiors as well as 
 chemical elements in upper MS stars (Maeder \& Meynet
\cite{Magn1}). In this work, we perform further theoretical developments
of the dynamo process, as well as
some numerical models,  to study the consequences of 
stellar evolution with a magnetic field.

 In the nice theory of  magnetic instabilities recently 
 developed by Spruit (\cite{Spruit02}), only some important
  limiting cases have been considered. In particular, the so--called Cases 0 and 1
  have been studied. In Case 0, the $\mu$--gradient dominates
over the thermal gradient in stellar interiors, i.e., in terms of the partial  Brunt--V\"{a}is\"{a}l\"{a}
frequency we have $N_{\mu} > N_{\mathrm{T}}$. Case 1 applies when 
the thermal gradient is the main restoring force, i.e.,
when $N_{\mu} < N_{\mathrm{T}}$. In this  case, the radiative losses
 from the magnetic instability are also accounted for; these losses reduce 
 the restoring force of buoyancy  which acts against the magnetic instability.
 
 The problem with the above description
  is that, in the  interiors of models no longer on the zero--age sequence, there is  at some moderate distance of the core
 a significant zone with $N_{\mu} < N_{\mathrm{T}}$ 
 (covering about 25 \% of the stellar radius), which is in a
 situation where the current  simplifications made in Case 1 are not appropriate. 
 In this zone (we called it zone 1P), the magnetic instability is adiabatic, while
 the hypothesis of large non-adiabaticity is made (see Sect.~2 below).
   It also happens in all stars   that there are zones
 where  $N_{\mu}$  and $N_{\mathrm{T}}$ are of the same order and where it is better to 
   account for both gradients, even if in general such zones are rather small.
  We consider it 
 preferable to  consistently treat the general case and to account  properly everywhere for
 the $T$-- and $\mu$--gradients, as well as for the non--adiabatic effects.
 
 Also, the two  approximations mentioned above
  have very different functional
 dependences with respect to stellar parameters, such as  the angular velocity
 $\Omega$, the differential rotation $q= -\frac{\partial \ln \Omega}{\partial \ln r}$
 or the thermal and chemical gradient. The use of asymptotic solutions such as
 Spruit's (\cite{Spruit02}) introduces  artificial jumps in
 the   behaviour of the diffusion
 coefficients.
 
 Our purpose here is to express  the theory in a unique and consistent formulation for 
 all  possible cases of  $N_{\mu}$ and  $N_{\mathrm{T}}$ and of departures from adiabaticity.
  In Sect. 2, we discuss the non--adiabaticity as a 
  function of radiative and magnetic diffusivities
  $K$ and $\eta$ respectively. In Sect. 3, we establish the general equations
  for the magnetic field  and  we check for  consistency
   with the expressions derived by Spruit (\cite{Spruit02}). In 
   Sect. 4 we examine the general expressions for the transport of 
   the angular momentum and of the chemical elements. Sect. 5 shows some initial numerical 
  applications  as an example. In Sect.~6 we examine the reciprocal
  feedback between meridional circulation and Tayler instability and the fact 
  that these two processes lead to very different amounts of differential rotation.
  The conclusions are given in Section 7.

\section{The radiative heat losses of the   magnetic instability}

\subsection{Magnetic diffusivity and radiative  losses }

As shown by Spruit (\cite{Spruit02}), the magnetic diffusivity $\eta$ in Case 0
($N_{\mu} > N_{\mathrm{T}}$) is given by

\begin{equation}
\eta_{\mathrm{0}} = r^2 \; \Omega \;\left(\frac{\omega_{\mathrm{A}}}{\Omega}\right)^4 
\left(\frac{\Omega}{N_{\mu}}\right)^2  \; ,
\label{eta0}
\end{equation}
\noindent
while in Case 1 ($N_{\mu} < N_{\mathrm{T}}$), it is given by

\begin{equation}
\eta_{\mathrm{1}} = r^2 \; \Omega \;\left(\frac{\omega_{\mathrm{A}}}{\Omega}\right)^2
\left(\frac{\Omega}{N_{\mathrm{T}}}\right)^{\frac{1}{2}}
\left(\frac{K}{r^2 N_{\mathrm{T}}}\right)^{\frac{1}{2}}  \; .
\label{eta1}
\end{equation}

\noindent
The Alfv\'en frequency $\omega_{\mathrm{A}}$ is given by

\begin{equation}
\omega_{\mathrm{A}} = \frac{B}{(4 \pi \rho)^{\frac{1}{2}} \; r} \; .
\end{equation}
\label{Alfven}
\noindent 
The Brunt--V\"{a}is\"{a}l\"{a}  frequency, when the $\mu$--gradient dominates over
the T--gradient, tends towards the following partial frequency
\begin{equation}
N_{\mu}^2 \; = \; \frac{g \varphi}{H_{\mathrm{P}}} \nabla_{\mu}  \; ,
\label{Nmu}
\end{equation}
\noindent
and when the T--gradient dominates, it tends towards

\begin{equation}
N_{\mathrm{T}}^{2}\; = \; \frac{g \delta}{H_{\mathrm{P}}}(\nabla_{\mathrm{ad}}-\nabla) \; .
\label{NT}
\end{equation}

\noindent
The thermodynamic coefficients $\delta$ and $\varphi$ are defined as follows:
$\delta= - \left(\frac{\partial \ln \rho}{\partial \ln T}\right)_{\mathrm{P},\mu}$ and
$\varphi=  \left(\frac{\partial \ln \rho}{\partial \ln \mu}\right)_{\mathrm{T,P}}$.
$H_{\mathrm{P}}$ is the pressure scale height.
We need a general expression for the magnetic diffusivity
encompassing the two limiting cases with $\eta_{\mathrm{0}}$ and
$\eta_{\mathrm{1}}$. This requires  further study of the nonadiabatic 
heat losses. For a fluid element displaced over a length $l$,
the thermal fluctuations diffuse with a timescale $t_{\mathrm{therm}}
\approx \frac{l^2}{  K}$, where $K$ is the thermal diffusivity
$K= \frac{4ac T^3}{3 \kappa \rho^2 C_{\mathrm{p}}}$.
Let us call $t$ the typical timescale over which the magnetic instability develops.
In a time $t$, the fluctuations are reduced
by a factor $f$, which can be written 
(cf. Spruit \cite{Spruit02}), 

\begin{equation}
f \;= \; \frac{t}{t_{\mathrm{therm}}} + 1   \; .
\label{f}
\end{equation}

\noindent
If $f \rightarrow 1$ as for  $t_{\mathrm{therm}} \gg t$, where t is the dynamical timescale of the
instability, one has the adiabatic case.
On the other hand, if $f \rightarrow \infty$ as for $t_{\mathrm{therm}} \ll t $, we have a highly 
nonadiabatic case. Spruit  (\cite{Spruit02}) also introduces in  Case 1
 (when $N_{\mathrm{T}}$ dominates) an ``effective thermal buoyancy
frequency'' $N_{\mathrm{e}}=\frac{N_{\mathrm{T}}}{f^{\frac{1}{2}}}$
and assumes $t_{\mathrm{therm}} \ll t$. This implies that $ f \simeq  
 \frac{t}{t_{\mathrm{therm}}}$. If it happens in some
 stellar  layer that 
 $t \ll t_{\mathrm{therm}}$, the situation is adiabatic and  one should  have  $f=1$, 
 but the adopted simplification leads to $f=0$. The approximation is thus invalid in this case.
 
 The factor $f$ introduced by Spruit
 is  related to the usual Peclet number $P_{\mathrm{e}} =
 \frac{t_{\mathrm{therm}}}{t_{\mathrm{dyn}}} \simeq
\frac{v \;l}{K}$, which is the ratio of the thermal to the dynamical
timescales  for a fluid element of velocity $v$ displaced over a length $l$. In astrophysics,
rather than the Peclet number,   one often considers (e.g. in the
theory of convection) a factor $\Gamma$, which is the ratio  between the  energy 
delivered by a fluid element to the energy lost during the displacement of the fluid element.
For a typical geometry, i.e., by assuming that the fluid elements are
spherical, one would have 
$\Gamma = \frac{1}{6} \; P_{\mathrm{e}}$. 
The fluid elements are likely not spherical in the relevant geometry 
for Tayler instability. We should rather consider thin slabs of 
thickness $l$, (we are indebted to Henk Spruit for this
remark).  If so, the relation between $\Gamma$ and the Peclet
number is  $\Gamma = \frac{1}{2} \; P_{\mathrm{e}}$.
Thus,  the following relation is valid for the factor $f$ introduced by Spruit
  and the usual number ratios $P_{\mathrm{e}}$ or $\Gamma$

\begin{equation}
\Gamma= \frac{1}{2} \; \frac{1}{(f-1)} \quad  \quad \mathrm{or}  \quad
f \;= \; \frac{1  \;+ \; 2 \; \Gamma}{2 \; \Gamma}  \; .
\label{Gammaf}
\end{equation}

\noindent
This comes directly from the abovementioned definition of the Peclet number
as well as from Eq.~(\ref{f}).
In the adiabatic case, $ f  \rightarrow 1$ and $\Gamma \rightarrow \infty$; 
if non--adiabatic effects are important, $f$ is large  and $\Gamma$ is very small. 

\subsection{Oscillations frequencies}

The general expression of the Brunt--V\"{a}is\"{a}l\"{a}
frequency  for a displaced fluid element is

\begin{equation}
N^2 = \frac{g \; \delta}{H_{\mathrm{P}}}(\nabla^{\prime} - \nabla + \frac{\varphi}{\delta}
\nabla_{\mu}) \; ,
\label{nabla}
\end{equation}

\noindent
where $\nabla^{\prime}$ is the internal gradient of the the fluid element,i.e., expressing how 
$T$ changes inside the fluid element during its motion, while
$\nabla$ is the T--gradient of the surrounding medium.
One also has the following relations between the various 
$T$--gradients (cf. Maeder \cite{Maeder95})

\begin{equation}
\nabla^{\prime} - \nabla = \frac{\Gamma}{\Gamma+1} (\nabla_{\mathrm{ad}} -\nabla)  \;
\label{nablaprime}
\end{equation}
\noindent
and
\begin{equation}
\nabla= \frac{\nabla_{\mathrm{rad}}+  \frac{2 \;\Gamma^2}{1+ \Gamma}\nabla_{\mathrm{ad}}}
{1+ \frac{2 \; \Gamma^2}{1+ \Gamma}}  \; ,
\label{nablarad}
\end{equation}

\noindent
with a numerical factor of 6 instead of 2, if we would have considered spherical fluid elements.
 Thus, in a radiative zone where there is some
motion of matter, which may be  non--adiabatic, the thermal gradient depends on the
efficiency  $\Gamma$ of the thermal transport  and  the gradient  may lie anywhere between the
adiabatic and radiative gradients depending on this efficiency.
If we write  $\Gamma$ in terms of  $f$, we have 

\begin{equation}
\frac{\Gamma}{\Gamma+1} = \frac{1}{2f-1} \; .
\label{gammaf}
\end{equation}

\noindent
Thus, we may also write the   Brunt--V\"{a}is\"{a}l\"{a} frequency  
taking account of radiative losses in terms of $f$, as follows

\begin{equation}
N^2 = \frac{1}{2f-1} N^2_{\mathrm{T}}+ N^2_{\mu}  \; .
\end{equation}

\noindent
When $N_{\mu}=0$, Spruit (\cite{Spruit02}) writes $N^2_{\mathrm{e}} = \frac{N^2_{\mathrm{T}}}{f}$,
which  has a dependence on $1/f$  for large values of $f$;
however we note that the numerical factors may introduce some differences.

We can go a step further in  expressing  $\Gamma$ in term of the ratio
$\eta / K$ of the magnetic to the thermal diffusivity. As seen above, one has 
  $\Gamma  \simeq \frac{1}{2} \; \frac {l^2}{t \; K}$, where $l$ is the length scale
  of the instability.
   Instabilities over smaller lengthscales 
  are removed by magnetic diffusivity. We consider here the marginal case  $l^2= \eta \; t$
  (cf. Spruit \cite{Spruit02}) and thus 
  we get
  
  \begin{equation}
  \Gamma = \frac{1}{2} \; \frac{\eta}{K}  \; .
  \label{getak}
  \end{equation} 
  
  \noindent
  Thus,  one obtains  the ratio  $\frac{\Gamma}{\Gamma+1} = \frac{\eta}
  {\eta \; + \; 2 K}$, which leads to the following expression for 
  the Brunt--V\"{a}is\"{a}l\"{a} frequency 
\begin{equation}
N^2 = \frac{\eta / K} {\eta / K  +2} \; N^2_{\mathrm{T}}+ N^2_{\mu}  \; .
\label{Nfinal}
\end{equation}

\noindent
This is the oscillation frequency  of a fluid element displaced
by Tayler--Spruit instability. The numerical factor of $2$ in the above expression is due to the geometry we considered;  in many cases this numerical factor  will
  introduce some minor differences 
with to Spruit (\cite{Spruit02}). The resulting differences will be rather small because this factor
will generally appear with a power smaller than $1$, as for example in Eq.~(\ref{omega11}).

\section{General equation for the magnetic diffusivity}

As emphasized by Spruit (\cite{Spruit02}), the energy of  the Tayler instability
(Tayler \cite{Tay73}) must be large enough to overcome the restoring force of 
buoyancy and this implies  that the Alfv\'en frequency must be larger than some limit.
Thus, the vertical extent $l_{\mathrm{r}}$ of the magnetic instability 
 must be limited  (cf. Spruit \cite{Spruit02}, Eq. 6),
 
 \begin{equation}
 l_{\mathrm{r}} <  \frac{r \; \omega_{\mathrm{A}}}{N}  \; .
 \label{lr}
 \end{equation} 

\noindent
However, if this  radial scalelength is too small, the perturbations will be 
quickly damped by the diffusion of the magnetic field characterized by $\eta$ and 
this implies $ l_{\mathrm{r}}^2 >  \frac{\eta}{\sigma_{\mathrm{B}}}$, 
where $\sigma_{\mathrm{B}}$ is the characteristic frequency of the magnetic field.
 In the absence of Coriolis force, this frequency is $\sigma_{\mathrm{B}}=\omega_{\mathrm{A}}$.
 However, as shown by Spruit (\cite{Spruit02}; see also Pitts \& Tayler \cite{Pitts86}),
 in a rotating star (typically when $\Omega > \omega_{\mathrm{A}}$) this frequency becomes  $\sigma_{\mathrm{B}} =
 (\omega_{\mathrm{A}}^2 / \Omega)$  due to the  Coriolis force. Therefore one has
 in this case

 \begin{equation}
 l_{\mathrm{r}}^2 >  \frac{\eta}{\sigma_{\mathrm{B}}}= 
 \frac{\eta \; \Omega} {\omega_{\mathrm{A}}^2}  \; .
 \label{lmin}
 \end{equation} 
 
 \noindent
The combination of these two limits together with Eq.~(\ref{Nfinal}) leads to the following
condition  $\left(\frac{\omega_{\mathrm{A}}}{\Omega}\right)^4  >  \frac{N^2}{\Omega^2} \;
\frac{\eta}{r^2 \; \Omega}$, where $N^2$ is given by Eq.~(\ref{Nfinal}).
For  the  marginal situation corresponding to equality, we have

\begin{equation}
\left(\frac{\omega_{\mathrm{A}}}{\Omega}\right)^4  =  \frac{N^2}{\Omega^2} \;
\frac{\eta}{r^2 \; \Omega}   \; .
\label{premier}
\end{equation} 

\noindent
This equation relates the two unknown quantities, the magnetic diffusivity 
 $\eta$ and the Alfv\'en frequency $\omega_{\mathrm{A}}$.
 
 For the particular  cases studied by Spruit (\cite{Spruit02}), 
 the result was a much simpler equation. 
Here, we find it preferable to establish  a second relation
between the two unknown quantities. The amplification time $\tau_{\mathrm{a}}$ of the 
Tayler--Spruit  instability, i.e., ``the timescale  on which
the radial field $B_{\mathrm{r}}$ is amplified  into an azimuthal field 
of the same order as the already existing azimuthal field'' 
can be written  $\tau_{\mathrm{a}}=  N /(\omega_{\mathrm{A}} \Omega q)$ (Spruit \cite{Spruit02}).
The equality of the two timescales  $\tau_{\mathrm{a}}$ and $\sigma_{\mathrm{B}}^{-1}$, according
to the expression  given above, leads to a second  equation  
$\frac{\omega_{\mathrm{A}}}{\Omega} = q \; \frac{\Omega}{N}$, as found by 
Spruit (\cite{Spruit02}; his Eq. 18). When account is taken of the expression of the 
Brunt--V\"{a}is\"{a}l\"{a}  (Eq.~\ref{Nfinal}),
this gives

\begin{equation}
\left(\frac{\omega_{\mathrm{A}}}{\Omega}\right)^2  =
\frac{\Omega^2 \; q^2}
{  N^2_{\mathrm{T}} \; \frac{\eta / K} {\eta / K  \;+ \; 2} + N^2_{\mu}}.
\label{deusse}
\end{equation} 

\noindent
The  above two equations (\ref{premier}, \ref{deusse}) form a coupled system
of degree 4 (see below) with  two unknown quantities,  
$\eta$ and $\omega_{\mathrm{A}}$. This system could be solved 
by standard procedures, however we first  make  some simplifications.
 We  eliminate the expression of $N^2$
between these two equations and obtain

\begin{equation}
\eta \;= \; \frac{r^2 \; \Omega}{q^2} \; \left( \frac{\omega_{\mathrm{A}}}
{\Omega}\right)^6  \; .
\label{eta}
\end{equation}

\noindent
This is a new general equation  connecting the magnetic diffusivity and the
Alfv\'en frequency. It shows that the Alfv\'en frequency, and therefore the magnetic field,
 changes as 1/6 the power of  the magnetic diffusivity for a given rotation.
This coefficient  applies to the diffusion of the 
azimuthal component of the magnetic field. Due to the particular pinch--type 
nature of the Tayler--Spruit instability, it  also 
applies to the turbulent diffusive mixing of the chemical elements by this instability.
To ensure consistency, 
we verify that in Spruit's Case 0 with 

\begin{equation}
\left (\frac{\omega_{\mathrm{A}}}{\Omega} \right )^2_0 = \left(
q \; \frac{\Omega}{N_{\mu}}\right)^2    \; ,
\label{omegaA0}
\end{equation} 

\noindent
we get 

\begin{equation}
\eta_0 = r^2 \Omega  q^4 \; \left(\frac{\Omega}{N_{\mu}}\right)^6  \; .
\label{etaz0}
\end{equation}
\noindent
This is in agreement with Eq.~(42) by Spruit (\cite{Spruit02}). 
This equation shows that the mixing of chemical elements decreases
very strongly if the $\mu$--gradient grows and this effect  limits
the chemical mixing of elements by the Tayler--Spruit dynamo in the regions
just above the convective core.

In Case 1, when
thermal losses are accounted for, the Alfv\'en frequency is given by

\begin{equation}
\left (\frac{\omega_{\mathrm{A}}}{\Omega}\right )_1 =  q^{\frac{1}{2}}
\left(\frac{\Omega}{N_{\mathrm{T}}}\right)^{\frac{1}{8}}
\left(\frac{K}{r^2 N_{\mathrm{T}}}\right)^{\frac{1}{8}}  \; 
\label{omegaA1}
\end{equation}

\noindent
This is given by Spruit (\cite{Spruit02}; his Eq.~19). Below, we will verify that 
our system of equations also leads to this expression (see Eq.~\ref{omega11} below).
For now,  we check that our  Eq.~(\ref{eta}) with this expression of 
$(\frac{\omega_{\mathrm{A}}}{\Omega} )_1$ leads to

\begin{equation}
\eta_1=    r^2 \Omega \; q
\left(\frac{\Omega}{N_{\mathrm{T}}}\right)^{\frac{3}{4}}
\left(\frac{K}{r^2 N_{\mathrm{T}}}\right)^{\frac{3}{4}}  \; ,
\label{etaz1}
\end{equation}

\noindent
which is the same as Eq.~(43) by Spruit (\cite{Spruit02}). Thus, we have checked the consistency 
of the general expression for $\eta$ (Eq.~\ref{eta}) with the results obtained by
Spruit (\cite{Spruit02}) in the two particular cases he considered. There are no new physics 
here with respect to Spruit's work (\cite{Spruit02}), we simply verify the consistency of our general expression with the asymptotic cases considered by Spruit.

Now, we can introduce the  general expression of $\eta$ given by
(\ref{eta}) in the second equation  (Eq.~{\ref{deusse}) of our coupled system and
obtain

\begin{eqnarray}
\left(\frac{\omega_{\mathrm{A}}}{\Omega}\right)^2 
\left[N_{\mathrm{T}}^2 \; \frac{r^2 \Omega}{q^2 K} 
\left(\frac{\omega_{\mathrm{A}}}{\Omega}\right)^6 +
N_{\mu}^2 \; \left (\frac{r^2 \Omega}{q^2 K}
\left(\frac{\omega_{\mathrm{A}}}{\Omega}\right)^6 +2\right )\right] = \nonumber \\[2mm]
\left( \frac{r^2 \Omega}{q^2 K} \;
 \left(\frac{\omega_{\mathrm{A}}}{\Omega}\right)^6 + 2\right) \; 
 \Omega^2 \; q^2. \quad \quad  \; 
\end{eqnarray}

\noindent
Since the ratio $\left(\frac{\omega_{\mathrm{A}}}{\Omega}\right)$ always appears 
with a power of 2, we now define a new variable 
 $x=\left(\frac{\omega_{\mathrm{A}}}{\Omega}\right)^2$ and thus  we get

\begin{equation}
\frac{r^2 \Omega}{q^2 K} \left(N_{\mathrm{T}}^2 + N_{\mu}^2 \right)  x^4-
\frac{r^2 \Omega^3}{K} x^3 + 2 N_{\mu}^2 \; x - 2 \Omega^2 q^2 = 0 \; .
\label{equx}
\end{equation}

\noindent 
We have transformed our system of two equations of degree 4 with two unknown quantities 
$\eta$ and $\omega_{\mathrm{A}}$
into one equation of degree 4  with only one unknown quantity $x$. The solution 
 will provide the value of the Alfv\'en frequency and the magnetic diffusivity
$\eta$ by Eq.~(\ref{eta}). Higher order 
terms depending on $\Gamma= \frac{1}{2} \frac{\eta}{K}$ might be considered, but as 
also shown by Spruit,
this ratio is very small. The above equation applies to the general 
case where both $\nabla_{\mu}$ and $N_{\mathrm{T}}$ are different from zero
and where thermal losses may reduce the restoring buoyancy force. This full equation 
is to be solved numerically (see Sect.~5 below). 

Higher order terms depending on $\Gamma^2$ could be considered, but it is correct  
to limit  the development to the first order in $\Gamma$.
Indeed, we notice that the expression of $N_{\mathrm{T}}$ given by Eq.~(\ref{NT})
contains the term $\nabla$, which is given by Eq.~(\ref{nablarad}).
This term $\nabla$ contains the 
 ratio $\eta \over K$. If we would fully develop the above equation taking this
 additional dependence into account we would get an equation of order 10
 instead of order 4 as above. We remark that the dependence
 in $\eta \over K$ introduced by  $\nabla$ in the expression of $N^2$ is of an higher order in
  $\eta \over K$ than the one expressed by Eq.~(\ref{Nfinal}).
   Since  $\eta \over K$ ranges between $10^{-2}$ and $10^{-7}$
   in the numerical models of
  Sect.~5, we may ignore this higher order term.

We can also check that Eq.~(\ref{equx}) leads 
to simple solutions fully consistent with the results by Spruit (\cite{Spruit02}) in 
the  particular cases  he has studied:\\

-- 1. Firstly, we notice that if terms going like 
the power 6 or higher of $\left(\frac{\omega_{\mathrm{A}}}{\Omega}\right)$
are negligible, the solution is simply

\begin{equation}
x = 
\left(q \; \frac{\Omega}{N_{\mu}}\right)^2    \; ,
\label{omzero}
\end{equation} 

\noindent
which is the same as Eq.~(\ref{omegaA0}) above. This is the so--called Case 0.
Indeed, the condition $\omega_{\mathrm{A}} \ll \Omega$ leads
to a solution which is the same as when $N_{\mathrm{T}}=0$, which defines Case 0.
\\

-- 2. The general Eq.~(\ref{equx}) can also be written

\begin{equation}
\frac{r^2 \Omega}{ K} x^3
\left( \frac{x}{q^2} (N_{\mathrm{T}}^2 + N_{\mu}^2) - \Omega^2 \right) 
+ 2  q^2 \left(\frac{x N_{\mu}^2}{q^2} - \Omega^2\right) = 0 \; .
\label{equxx}
\end{equation}

\noindent
We notice some similarity of the two terms in parentheses. If we consider
Spruit's Case 0 with $N_{\mathrm{T}}=0$, this equation becomes the product
of two terms

\begin{equation}
\left( \frac{x}{q^2} N_{\mu}^2 - \Omega^2 \right) 
  \left( \frac{r^2 \Omega}{ K} x^3 + 2 q^2\right) = 0 \; .
\label{equxxx}
\end{equation}

\noindent
The solutions are

\begin{equation}
x= \frac{q^2 \Omega^2}{N_{\mu}^2} \quad \quad \mathrm{and} \quad 
x^3= - \frac{2 \; q^2 K}{r^2 \Omega} \; .
\end{equation}

\noindent
The second possibility leads to negative solutions, which are thus 
not physically meaningful. Thus, this shows the uniqueness of the
first solution.This first solution is that of Case 0,
given above by  Eq.~(\ref{omegaA0}), as found by Spruit (\cite{Spruit02}).\\

-- 3. Let us consider Case 1 with $N_{\mu}=0$. We have from Eq.~(\ref{equxx})

\begin{equation}
\frac{r^2 \Omega}{ K} x^3
\left( \frac{x}{q^2} N_{\mathrm{T}}^2  - \Omega^2 \right) 
- 2  \;  q^2 \Omega^2 = 0 \; .
\end{equation}

\noindent 
If $q$ is small, one has the solution 

\begin{equation}
 x = \frac{q^2 \Omega^2}{N^{2}_{\mathrm{T}}}  \; ,
 \label{1P}
 \end{equation}
 
\noindent 
which is the alternative solution proposed by Spruit(\cite{Spruit02}), when the
simplification resulting from the assumption of a large $f$ value does not apply,
(this is the case we have called 1P).

We may also examine the general solution when the ratio $\eta / K$ is small, which 
 is generally the case as seen above and as mentioned by Spruit.
   For that, we start from the two equations
(\ref{premier}, \ref{deusse}) and make the simplifications appropriate 
for a small  $\eta / K$  ratio. With  developments
like those done above, we get instead of Eq.~(\ref{equx}) the following expression

\begin{equation}
\frac{r^2 \Omega}{q^2 K} N_{\mathrm{T}}^2  \; x^4
 + 2 N_{\mu}^2 \; x - 2 \; \Omega^2 q^2 = 0 \; ,
\label{equetaK}
\end{equation}

\noindent
which is similar to our Eq.~(\ref{equx}) for  $N_{\mu}=0$, except that the 
term in  $x^3$ has disappeared. Indeed, we may also derive the
 above equation directly from our general  Eq.~(\ref{equx}) by noting that if 
 $\eta/K$ is small, then from Eq.~(\ref{deusse}) one has 
 
\begin{equation}
x= \frac{\Omega^2 \; q^2}{N^2_{\mu}}
\end{equation}

\noindent
and then by multiplying this equation by $\frac{x^3 r^2 \Omega N^2_{\mu}}
{q^2 K}$, we get

\begin{equation}
x^4 \frac{r^2 \; \Omega}{q^2 \; K} N_{\mu} - \frac{r^2\; \Omega^3}{K}\; x^3 =0  \; .
\end{equation}

\noindent
When we bring this relation into Eq.~(\ref{equx}), we are  lead to the above 
Eq.~(\ref{equetaK}). Thus, this relation (\ref{equetaK})
is fully consistent with our general equation.

 In Case 1 when $N_{\mu}$, the above
 Eq.~(\ref{equetaK}) leads to 

\begin{eqnarray}
\left(\frac{\omega_{\mathrm{A}}}{\Omega}\right)=
2^{\frac{1}{8}} \;q^{\frac{1}{2}} 
\left(\frac{\Omega}{N_{\mathrm{T}}}\right)^{\frac{1}{4}}
\left( \frac{K}{r^2 \Omega}\right)^{\frac{1}{8}} =   \nonumber \\[2mm]
2^{\frac{1}{8}} \;q^{\frac{1}{2}} 
\left(\frac{\Omega}{N_{\mathrm{T}}}\right)^{\frac{1}{8}}
\left( \frac{K}{r^2 N_{\mathrm{T}}}\right)^{\frac{1}{8}} \; .
 \label{omega11}
\end {eqnarray}

\noindent
This is just the same expression, except
for the small geometrical factor $2^{\frac{1}{8}} $, as in the equation
 found by Spruit (\cite{Spruit02}, i.e.,his Eq. 19). 

Thus, for all situations considered
by Spruit,  we have  verified  here that our 
general equations (\ref{eta}, \ref{equx}) well reproduce his results. 
 
 \section{Transport of angular momentum and chemical elements}
 
 Let us specify some other  useful expressions such as the critical 
 lengthscale for the Spruit--Tayler instability, the intensity of the magnetic field,
 the coefficients of transport for the angular momentum and for the chemical elements.
 As shown above, the vertical extent of the magnetic instability is limited by
 $l_{\mathrm{r}} <  \frac{r \; \omega_{\mathrm{A}}}{N} $ (Eq.~\ref{lr}),
 where N is given by the general expression (\ref{Nfinal}). Thus the  maximum
 lengthscale of the magnetic instability is
 
 \begin{equation}
 l_r= \frac{r \; \omega_{\mathrm{A}}}
 {\left( \frac{\eta / K} {\eta / K  +2} \; N^2_{\mathrm{T}}+ N^2_{\mu}\right)
 ^{\frac{1}{2}}}  \; .
 \label{lgeneral}
 \end{equation}
 
 \noindent
 Following Spruit (\cite{Spruit02}), we consider this maximal value as the one
 charateristic, for example, of mixing, since the mixing will be essentially 
 determined by  the  instabilities with the longest lengthscale.
 We immediately verify that $l_r$ is in agreement with Spruit's (\cite{Spruit02})
  result in Case 0, when $N_{\mathrm{T}}=0$.
 In Case 1, when $N_{\mu}=0$, we may write,  if we assume in addition that $\frac{\eta}{K}$ is small,
 
 \begin{equation}
  l_r= \frac{r \; \omega_{\mathrm{A}}}
  {\left(\frac{\eta}{2 K}\right)^{\frac{1}{2}} N_{\mathrm{T}}} \: .
 \end{equation}
 \noindent
 With the  general  expression of $\eta$ given by Eq.~(\ref{eta}), this becomes
 
 \begin{equation}
 l_{\mathrm{r}}= \frac{2^{\frac{1}{2}} \; K^{\frac{1}{2}} \; q \; \Omega^{\frac{5}{2}}}
 {N_{\mathrm{T}} \; \omega_{\mathrm{A}}^2}  \; .
 \end{equation}
 
 \noindent
 With Eq.~(\ref{omega11}) for $\omega_{\mathrm{A}}$, we get finally in Case 1
 \begin{equation}
 l_{\mathrm{r}}=2^{\frac{1}{4}} \; r \; \left(\frac{\Omega}{N_{\mathrm{T}}}\right)
 ^{\frac{1}{2}}
 \left(\frac{K}{r^2 \Omega}\right)^{\frac{1}{4}} \;,
 \label{lmax1}
 \end{equation}
 
 \noindent
 which is identical to the result by Spruit (\cite{Spruit02}; Eq.~10), apart
 from the  geometrical factor $2^{\frac{1}{4}}$. Thus, our general expression
 for the  lengthscale $l_{\mathrm{r}}$ of the Tayler--Spruit instability 
 also contains the limiting cases studied by Spruit. We also recall, as shown 
 in Eq.~(12) by Spruit, that $\eta \ll K$ is a condition of validity for the developments
 made by Spruit.
 
 Turning now to the general expression  for the intensity of the magnetic field, 
 following Spruit (\cite{Spruit02}) we have for the azimuthal and radial field strengths,
 
 \begin{equation}
  B_{\varphi}= (4 \pi \rho)^{\frac{1}{2}} \; r \; \omega_{\mathrm{A}} \quad \mathrm{and} \quad
 B_ {\mathrm{r}}= B_{\varphi} \; (l_{\mathrm{r}} / r)  \; .
 \label{champ}
 \end{equation}
 
 \noindent
 The quantity
 $\omega_{\mathrm{A}}$  has  to be taken as the solution of the general
 equation (\ref{equx}) and $l_{\mathrm{r}}$ is given by Eq.~(\ref{lgeneral}).
  These general  expressions  also reproduce correctly the particular
 Cases 0 and 1,
 since we have verified the consistency of our general expressions  for $l_{\mathrm{r}}$
 and $\omega_{\mathrm{A}}$ with those for the two particular cases.
 For Case 0, this gives (cf. Spruit (\cite{Spruit02}),
 
\begin{equation}
B_{\varphi}= (4 \pi \rho)^{\frac{1}{2}} r \; q \frac{\Omega^2}{N_{\mu}}  \; ,
\quad \mathrm{and} \quad
\frac{B_ {\mathrm{r}}}{B_{\varphi}}  = \; q 
 \left(\frac{\Omega}{N_{\mu}} \right)^2  \; .
\label{Bphi}
\end{equation}

 \noindent
 For Case 1, one has
 
 \begin{equation}
B_{\varphi}=  2^{\frac{1}{8}} (4 \pi \rho)^{\frac{1}{2}} \; r \; \Omega  \; q^{\frac{1}{2}}
 \left(\frac{\Omega}{N_{\mathrm{T}}}\right)^{\frac{1}{8}}
 \left(\frac{K}{r^2 \; N_{\mathrm{T}}}\right)^{\frac{1}{8}}
 \end{equation}
 \noindent
 and
 \begin{equation}
 \frac{B_ {\mathrm{r}}}{B_{\varphi}}  =2^{\frac{1}{4 }}
 \left(\frac{\Omega}{N_{\mathrm{T}}}\right)^{\frac{1}{4}}
 \left(\frac{K}{r^2 \; N_{\mathrm{T}}}\right)^{\frac{1}{4}}  \; .
 \end{equation}
 \noindent
 These last two expressions are identical to those by  Spruit
 except for the geometrical factors  $2^{\frac{1}{8}}$ and 
 $2^{\frac{1}{4}}$ respectively.
 
 Let us now turn towards the transport of angular momentum by the magnetic field.
  The azimuthal stress  by volume unity due to the magnetic field  is given by
  
  \begin{eqnarray}
  S \; = \frac{1}{4 \; \pi} \; B_{\mathrm{r}} B_{\varphi} \; = \;
  \frac{1}{4 \; \pi} \;  \left(\frac{l_{\mathrm{r}}}
  {r}\right) B_{\varphi}^2 = 
  \; \rho \; r^2 \; \left(\frac{\omega_{\mathrm{A}}^3}{N}\right)  \; .
  \end{eqnarray}

 \noindent
 Following the ingenious procedure devised by Spruit (\cite{Spruit02}), we can 
 express the viscosity $\nu$
 for the vertical transport of angular momentum in terms of  $S$.  Indeed, the viscosity
 of a fluid represents its ability to transport momentum from one place to another 
 and thus we get,
 
 \begin{equation}
 \nu = \frac{S}{\rho \; q \; \Omega} =
  \; \frac{\Omega \; r^2}{q} \;
 \left( \frac{\omega_{\mathrm{A}}}{\Omega}\right)^3 \; 
\left(\frac{\Omega}{N}\right) \; .
 \label{nu}
 \end{equation}
 
 \noindent
 This is the general expression of $\nu$ with $\omega_{\mathrm{A}}$ given by the solution
 of Eq.~(\ref{equx}) and with $N$ by Eq.~(\ref{Nfinal}). We immediately  verify that in Case 0
 with $\frac{\omega_{\mathrm{A}}}{\Omega} = \; q   \; \frac{\Omega}{N_{\mu}}$,
 we find the same as Spruit (\cite{Spruit02}),
 
 \begin{equation}
 \nu_0 = r^2 \; \Omega \; q^2 \; \left( \frac{\Omega}{N_{\mu}} \right)^4 \; .
 \label{nuzero}
 \end{equation}
 
 \noindent
 The $\mu$--gradient, through its reduction of the field, particularly in the vertical direction,
  also reduces the transport of angular momentum, but
 much less than for the chemical elements as given by Eq.~(\ref{etaz0}).
 In Case 1, it is also easy to verify, by using the appropriate expressions  for N,
  $l_{\mathrm{r}}$ and  $\omega_{\mathrm{A}}$ that one obtains,
 
\begin{equation}
 \nu_	1= 2^{\frac{1}{2}} r^2 \; \Omega \; \left(\frac{\Omega}{N_{\mathrm{T}}}\right)
 ^{\frac{1}{2}} \left(\frac{K}{r^2 N_{\mathrm{T}}}\right)^{\frac{1}{2}} \; ,
 \label{nu1}
 \end{equation}
 
 \noindent
 which is identical to Eq.~(32) by Spruit (\cite{Spruit02}) apart from the geometrical
 factor $6^{\frac{1}{2}}$.
 
 Thus, we have the full set of expressions necessary  to obtain  
 the Alfv\'en frequency $\omega_{\mathrm{A}}$, the magnetic diffusivity  $\eta$, which is necessary for calculating
 the transport of the chemical elements, and the viscosity $\nu$
  for the vertical transport   of the angular momentum by the magnetic field. These
  expressions apply to the general case, where both the $\mu$-- and $T$--gradients 
  are different from zero and where radiative losses may play a role. 
  In all cases, the general expressions may also lead consistently to the
  asymptotic expressions by Spruit (\cite{Spruit02}).
  We may remark that
  we need not to include here the erosion of the $\mu$--gradient by the horizontal
  turbulence as suggested by Talon \& Zahn (\cite{TalonZ}), since as shown by Maeder \& Meynet
  (\cite{Magn1}), the horizontal turbulence is either suppressed or at least strongly reduced
  by the magnetic field.

   \begin{figure}[t]
  \resizebox{\hsize}{!}{\includegraphics{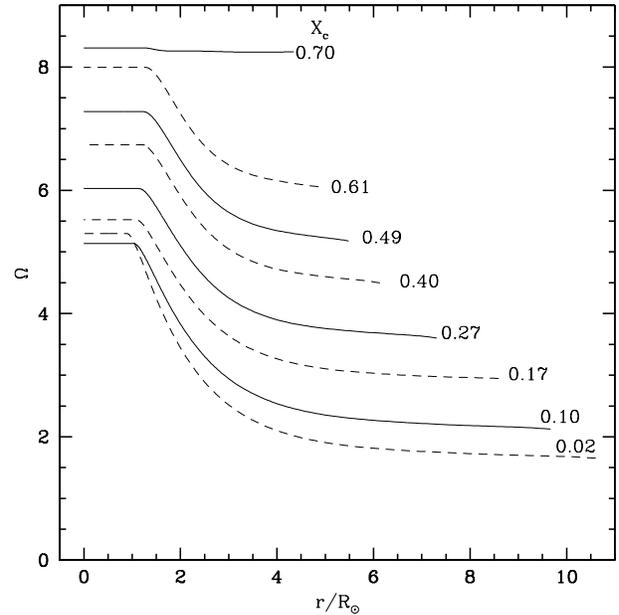}}
  \caption{Internal distribution of the angular velocity $\Omega(r)$ as a function of the radius in solar units in the model  V3 (without magnetic field) 
  at various stages of the model 
  evolution indicated by the central H--content $X_{\mathrm{c}}$ during the MS--phase.
  The initial mass is
  15 M$_{\odot}$ and  $Z=0.02$.}
\label{omegav3}
\end{figure}

 \begin{figure}[t]
  \resizebox{\hsize}{!}{\includegraphics{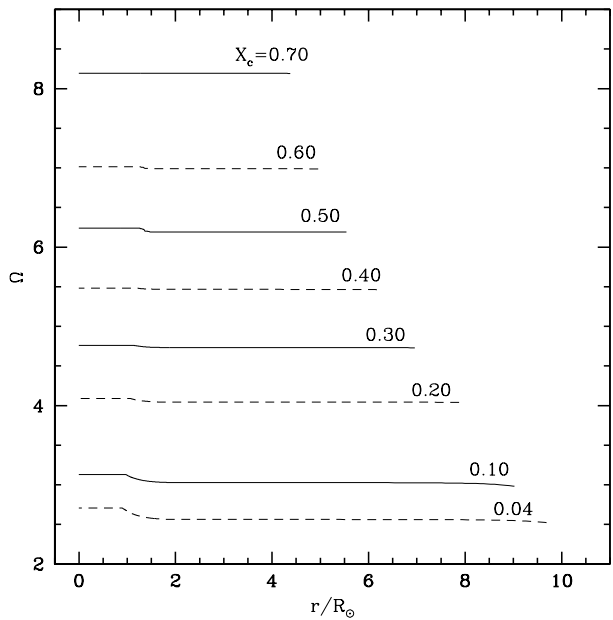}}
  \caption{ Internal distribution of the angular velocity $\Omega(r)$ as a function of the radius in solar units  in the model  M1 (with magnetic field calculated according to
  Spruit \cite{Spruit02}) at various stages of the model 
  evolution indicated by the central H--content $X_{\mathrm{c}}$ during the MS--phase.
  The initial mass is
  15 M$_{\odot}$ and $Z=0.02$.}
\label{omegaM1}
\end{figure}

  Finally, let us check consistency. The rate of magnetic energy production 
  $W_{\mathrm{B}}$ per unit of time and volume must be equal to the rate
  $W_{\nu}$ of the dissipation of rotational energy by the magnetic viscosity
  $\nu$ as given above. This check of consistency is verified for the asymptotic
  expressions, but we have to see whether the general expressions also fulfill it.
  We assume here   that the whole energy 
  dissipated is converted to magnetic energy, which is  a reasonable approximation.
   The sink of energy due to chemical mixing is  negligible
  here, since as shown below the transport of elements from the core to the surface is absent from  models with the magnetic field as calculated here.
  
  \begin{equation}
W_\nu={1\over 2}\rho\nu \Omega^2q^2 ,
\label{Wnu}
\end{equation}
\noindent
which gives with Eq.~(\ref{nu}),

\begin{equation}
W_\nu={1\over 2}\rho \; q \; r^2 \frac{\Omega}{N} \omega_{\mathrm{A}}^3  \; .
\label{nunu}
\end{equation}

\noindent
With Eq.~(\ref{deusse}) defining the field amplitude  $\left(\frac{\omega_{\mathrm{A}}}{\Omega}\right)= 
\frac{\Omega \;  q}{N}$, the dissipation rate finally becomes

\begin{equation}
W_\nu={1\over 2}\rho  r^2 q^4 \Omega^3 \left(\frac{\Omega}{N}\right)^4  \; .
\label{nununu}
\end {equation}

\noindent
We turn to the rate $W_{\mathrm{B}}$ of magnetic energy creation.
The magnetic energy per unit volume is $\frac{B^2}{8 \pi}$, it
is produced in a characteristic time given by $\sigma_{\mathrm{B}}^{-1} =
 (\omega_{\mathrm{A}}^2 / \Omega)^{-1}$. Thus, one has

\begin{equation}
W_{\mathrm{B}}= \frac{B^2}{8 \pi} \frac{\omega_{\mathrm{A}}^2}{\Omega}=
\frac{1}{2} \rho r^2 \frac{\omega_{\mathrm{A}}^4}{\Omega}  \; ,
\label{WB}
\end{equation}

\noindent
where we have used the Eq.~(\ref{champ}) for the field, because
$B_{\varphi}$ is the main field component.
If we now use the above expression of $\frac{\omega_{\mathrm{A}}}{\Omega}$,
we  get the same expression as for $W_{\nu}$ (Eq.~\ref{nununu}), thus one has

\begin{equation}
W_{\nu} \; = \; W_{\mathrm{B}}.
\end{equation}

\noindent
This shows the consistency
of the field expression for $B_{\varphi}$, of the transport coefficient
$\nu$ together with the energy conservation in the process of field creation.

 \section{Numerical applications}
 
 \subsection{Existing models, present models and their aims}

  \begin{figure}[t]
  \resizebox{\hsize}{!}{\includegraphics{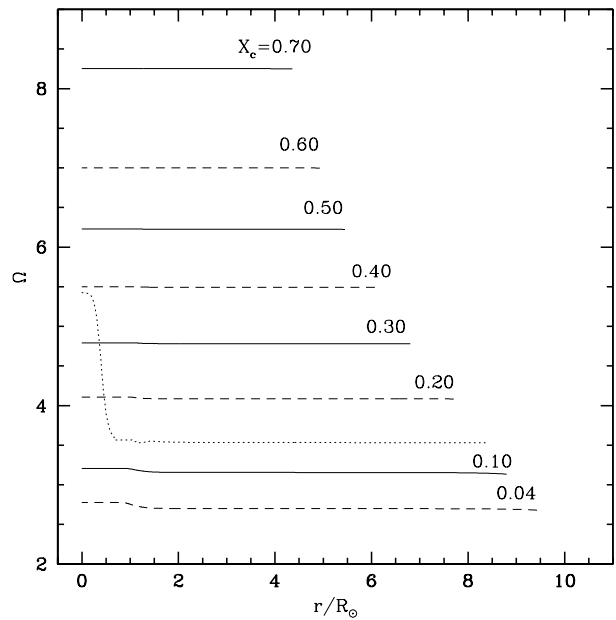}}
  \caption{Internal distribution of the angular velocity $\Omega(r)$ as a function of the radius in the model  M5 (with magnetic field calculated according to
  expressions of the present paper) at various stages of the model 
  evolution indicated by the central H--content $X_{\mathrm{c}}$ during the MS phase. The 
  dotted line shows the values at the end of the H--burning phase.}
\label{omegaM5}
\end{figure}

Many sets of models with rotation exist,  often without  magnetic fields.
A general review of the models without magnetic field has been made by 
Maeder \& Meynet (\cite{MMaraa}) and more recent references 
are given in the recent IAU Symposium 215 on Stellar Rotation 
(Maeder \& Eenens  \cite{MEe}). Most models cover the H-- and He--burning phases,
while some models go up to the pre--supernova stage (Heger et al. \cite{HegerI}; Hirschi
et al. \cite{Hirschi04}). The models often differ significantly in their
physical assumptions and the way they are modelized. A particularly critical point is
the treatment of the meridional circulation, which in many sets of recent models 
is treated as a diffusion process, which is incorrect since a diffusion process
goes along the gradient of the considered quantity (for example $\Omega$),
 while the advection due to meridional
circulation may transport the angular momentum from the location where 
there is less (central regions) to the location where there is more 
(superficial regions), and it can also do the opposite depending
on the circulation patterns.

Several stellar models with rotation and 
magnetic fields have been published in the recent literature.
The interaction between meridional circulation and dynamo action have been studied
by Charbonneau \& MacGregor (\cite{Charb01}; see also Sect.~6 below). 
Heger et al. (\cite{Heger04}) have implemented Spruit's (\cite{Spruit02})
 asymptotic developments in models of stellar evolution.  These authors follow  from
the ZAMS to the pre-supernovae stage  the evolution of the
specific angular momentum. They find in particular that the models with magnetic field 
lead to neutron stars spinning about an order of magnitude slower
 than the  models without magnetic field by Heger et al. (\cite{HegerI}.
  The model with magnetic field  is still rotating faster  than 
observed in young pulsars, while it is
 to slow for making the collapsar model possible
  (Woosley \cite{Woosley93}; MacFadyen et al. \cite{MacF01}).
The effect of magnetic field in the core collapse have been studied by Fryer \& Warren
(\cite{FryerW04}), who find that magnetic fields do not play a dominant role in the 
supernova explosion mechanism.
 
 In paper I of this series (Maeder \& Meynet \cite{Magn1}), we have
  calculated  a model of a 15
 M$_{\odot}$ with solar composition and  an initial rotation velocity
   $v_{\mathrm{ini}}$ of 300 km s$^{-1}$. At a
 stage near the middle of the MS phase, we have examined what happens if we
 ``turn on'' the magnetic field  and apply Spruit's expressions
  (Spruit \cite{Spruit02}). The general result was that both the  transports 
  of angular momentum and of chemical elements by 
 Tayler magnetic instability was much larger than the transports by meridional
 circulation  and shear instability. Thus, this leads us to suspect that magnetic field 
 is an important component of stellar evolution.
 \begin{figure}[t]
  \resizebox{\hsize}{!}{\includegraphics{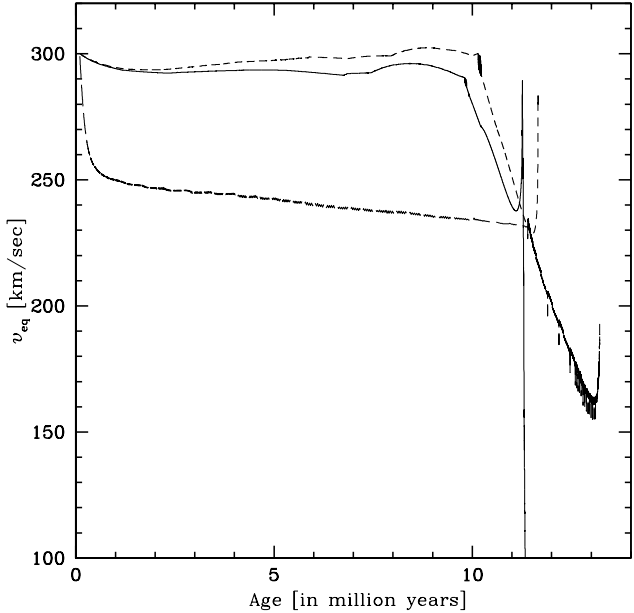}}
  \caption{ Evolution of the rotation velocities at the surface of the various models 
  with an initial mass of
  15 M$_{\odot}$ with $Z=0.02$,
  during the MS phase. The heavy broken line is for model V3  the light
  broken--line for model M1 and the continuous line for model M5.}
\label{rotationsurf}
\end{figure}
 Here, we want to compare Spruit's solutions and the more 
  general developments expressions developed 
 here. Also, we want  to study the effects of the field 
 through the evolution.
  For that purpose, we follow the MS evolution of 4 different types of models of 15 M$_{\odot}$ 
 with  solar composition. These  models are labeled as follows:
 \begin{itemize}
 \item V0: a model with no rotation and no magnetic field.
 \item V3: a model with $v_{\mathrm{ini}} = 300$  km$\cdot$s$^{-1}$, 
 and no magnetic field, with the physical assumptions adopted by  Meynet \& Maeder (\cite{MMX}) except
 for the overshooting parameter which was taken here equal to 0.
 \item M1: a model with $v_{\mathrm{ini}}$ of 300 km$\cdot$s$^{-1}$ and 
 with magnetic field calculated 
 according to the expressions given by Spruit (\cite{Spruit02}), i.e., for Case 0 
 when  $N_{\mu} > N_{\mathrm{T}}$ and  for Case 1 when  $N_{\mu} < N_{\mathrm{T}}$.
 In Case 1, when the account of radiative losses does not increase the magnetic
 diffusivity, another case called 1P as given by Eq.~(\ref{1P}) is considered
    (Spruit \cite{Spruit02}).
  \item M5: a model with $v_{\mathrm{ini}}$ of 300 km$\cdot$s$^{-1}$ and with magnetic field
 according to the expressions of the present paper.
 \end{itemize}
 
 \noindent
 In both models M1 and M5, we did not include  the effects of meridional circulation
 and  horizontal turbulent diffusion, since these effects were
 found small with respect to those due to the magnetic field, (see remarks in Sect.~6).
 In these first basic tests, overshooting and  semiconvection 
 were not included.
  However, we keep
 the effects of shear mixing and
 the hydrostatic effects of rotation, which distort the equipotentials and 
 modify the stellar shape, which has great consequences on the $T_{\mathrm{eff}}$
 of rotating models.  To make the comparisons more relevant, we do not include  in model M5
  the energy condition  discussed in paper I, since this condition was not
  included by Spruit (\cite{Spruit02}). Some further  
 theoretical works are still needed about this condition  and this point
  will be examined in a further work.

\subsection{Evolution of the internal and superficial rotation}

We examine the evolution of the internal profile of $\Omega(r)$ in the 
various models. Figs.~\ref{omegav3}, \ref{omegaM1} and \ref{omegaM5} show these
profiles at various evolutionary stages during the evolution of models
 V3, M1 and M5 respectively.
In model V3, we notice the smooth, but significant decline of $\Omega(r)$
outside the convective core, where rotation is  homogeneous. Such an
evolution of rotation has been studied in detail by  Meynet \& Maeder (\cite{MMV}) 
 and Maeder \& Meynet (\cite{MMVII}) at solar composition and at lower metallicity. 
 The angular velocity is first decreasing in the core as mass loss removes 
 angular momentum at stellar surface, because  enough coupling of rotation 
 is ensured by shear mixing and meridional circulation 
 to largely compensate the acceleration of rotation which would result from
 the moderate core contraction.
 Only at the end of the MS 
 phase, central rotation increases due to the dominant effect of central contraction. 
 In the outer layers, a smooth but significant decrease of rotation
 between the core and the surface  is created by the abovementioned processes.

 \begin{figure}[tb]
  \resizebox{\hsize}{!}{\includegraphics{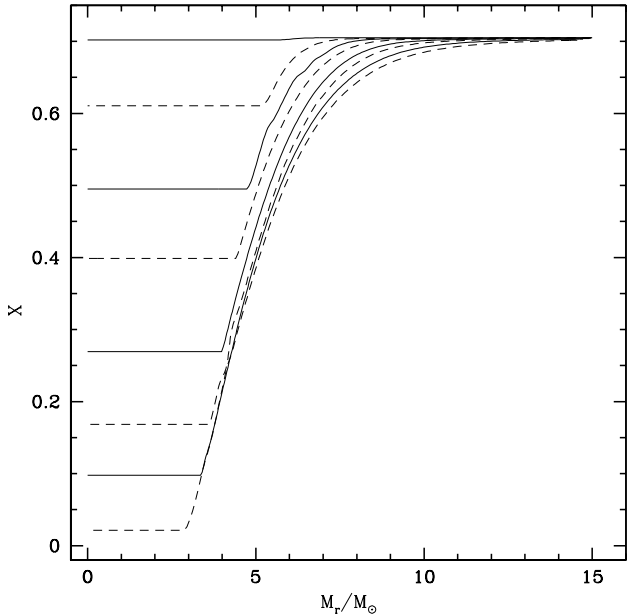}}
  \caption{ Internal distribution of the hydrogen mass fraction $X$
   as a function of the Lagrangian mass in the model  V3 (with rotation only)
    at various stages of the model  evolution from top to bottom during the MS--phase.}
\label{xE0}
\end{figure}

 \begin{figure}[tb]
  \resizebox{\hsize}{!}{\includegraphics{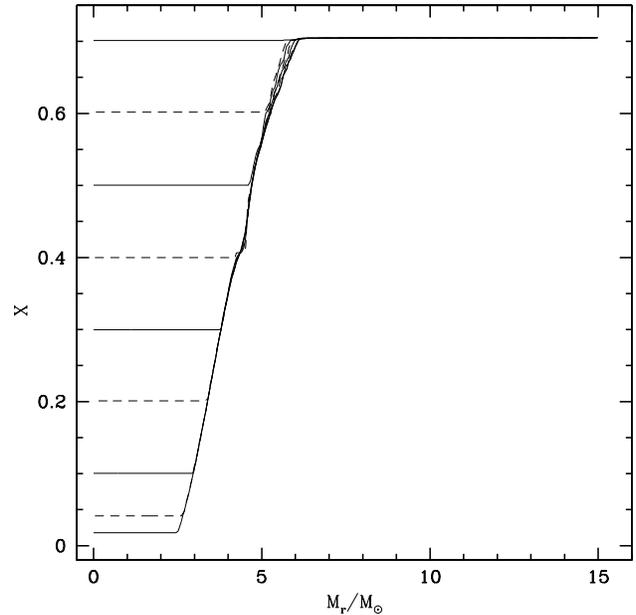}}
  \caption{Internal distribution of the hydrogen mass fraction $X$
  as a function of the Lagrangian mass in the model  M5 (with magnetic field calculated according to
  expressions of the present paper) at various stages of the model 
  evolution from top to bottom during the MS--phase.}
\label{xM5}
\end{figure}

 In Fig.~\ref{omegaM1} and \ref{omegaM5}, we notice several interesting new facts
 for models with a magnetic field:
 \begin{itemize}
 \item The models have  almost an internal solid body rotation, with a difference of only
 a few percent between the core and the bulk of the envelope, with a short
 transition at their interface. The differential rotation parameter $q$ is
 very small, but still different from zero.  The relative constancy of $\Omega$
 is the result of the high value of the transport 
 coefficient $\nu$ by the magnetic field (see below Fig.~\ref{coeffdiff}). We notice  that
 these results are in agreement with those of  Heger
 et al. (\cite{Heger04}), who also find essentially solid body rotation until the end 
 of the MS phase. Then in later phases, differential rotation appears together with
 a slowing down of the core. 
 \item There is a general slowing down of  rotation during evolution due to two
  effects: the mass loss which removes angular momentum and
 the expansion of the
  external layers. The strong internal coupling
  of rotation transports this slowing down to the core.
  \item Models M1 and M5, although not strictly identical (see Fig.~\ref{rotationsurf}), exhibit
 similar $\Omega$--distributions.
  \item We notice a small decrease of $\Omega$
   near the surface. This is likely resulting from   the  expansion
    of the  superficial layers after the removal of some mass by stellar winds at the surface.
   \end{itemize}
   
   Fig.~\ref{rotationsurf} shows the evolution of the surface velocities 
   for models V3, M1 and M5. We note the
   small differences between models  M1 and M5, showing that the general treatment
   is not identical to Spruit's solutions.
   We see that the fast initial decline  
   of model V3, due to the rapid adjustment of an equilibrium profile
   by meridional circulation  (cf. Meynet \& Maeder \cite{MMV})
   is absent in models with magnetic field which rotate nearly like a solid body
   from the ZAMS onwards.
   Thus models M1 and M5 keep an almost constant surface velocity, while model V3 also
   has a slight decline during the rest of the MS phase.
   Very likely this behaviour is different for different initial
   masses, because of the differences in the mass loss rates.
   Thus, a complete grid has to be made in order to establish a
   basis for comparisons with the observations. 

\subsection{Internal distribution of hydrogen}

 The evolution of the internal distribution of hydrogen mass fraction is illustrated 
 in Fig.~\ref{xE0} for model V3 and in Fig.~\ref{xM5} for  model M5
 with magnetic field, (models M1 and M5  gives very  similar curves with only
  minor differences as illustrated by the evolution of the core sizes in 
  Fig.~\ref{Mcc}).
 We see that at the end of the MS phase model M5 produces a  convective
 core which  is slightly smaller than for model V3. The H--distribution
  outside the core 
 is much smoother in model V3: there is more helium outside the formal core 
 of this model than in model M5. The reason is that in the magnetic models 
  the transport of chemical elements is  strongly 
  inhibited by the  $\mu$--gradient just outside the core: the diffusion 
  coefficient  for chemical elements depends on   $N_{\mu}$
  with a power  $-6$, as shown by  Eq.~(\ref{etaz0}). 
 In the models with only rotation, the inhibition by $N_{\mu}$ for the shear diffusion
 goes like a power $-2$.
 The fact that more helium is  burnt in models V3 has some consequences for the shape of the
 tracks in the HR diagram and also for the lifetimes as shown in Sect.~5.6 below.

\begin{figure}[t]
  \resizebox{\hsize}{!}{\includegraphics{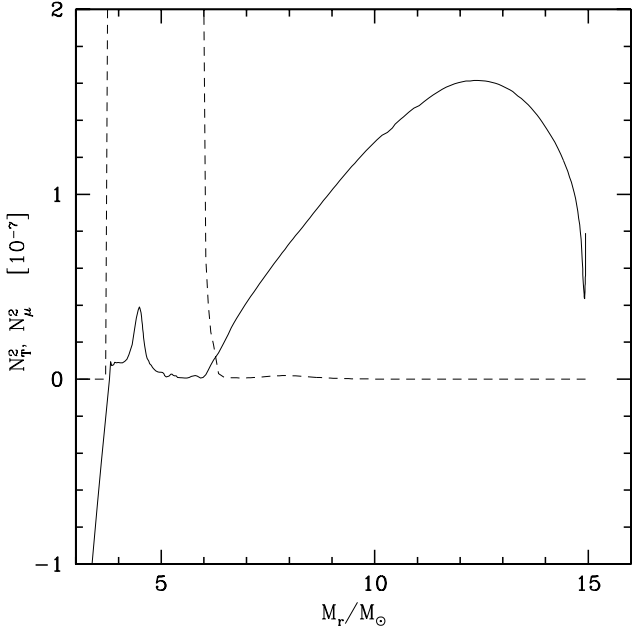}}
  \caption{Internal distribution of the partial Brunt--V\"{a}is\"{a}l\"{a} frequencies $N^2_{\mathrm{T}}$ and $N^2_{\mu}$ 
 for the model M5 
at an age of $8.384 \cdot  10^6$ yr with central H--content given by $X_{\mathrm{c}}= 0.30 $.
}
\label{NM5}
\end{figure}

\subsection{The Brunt--V\"{a}is\"{a}l\"{a} and Alfv\'en frequencies}

Fig.~\ref{NM5} shows the terms $N^2_{\mathrm{T}}$ and $N^2_{\mu}$ of the 
Brunt--V\"{a}is\"{a}l\"{a} frequency for the model M5 near the middle of the
MS phase, (the general behaviour is similar for model
M1, thus we do  not show it). We see that the thermal component 
$N^2_{\mathrm{T}}$ dominates  through most of the envelope while the mean
molecular weight component $N^2_{\mu}$ is much more important close to the core. 
This large $N^2_{\mu}$  inhibits the chemical mixing.

 The ratio $\omega_{\mathrm{A}}/ \Omega$
lies  everywhere between $10^{-3}$ and $10^{-2}$.
The magnetic field  $B_{\varphi}$ is nearly constant through the star.
Its amplitude is  about  $\log B_{\varphi}= 4.4$ throughout with average deviations smaller
than about 0.1 dex, where the field $B_{\varphi}$ is expressed in Gauss.
 As already suggested by 
Maeder \& Meynet (\cite{Magn1}), the field 
currently due to  Tayler--Spruit dynamo is rather large in massive stars.
The situation is different here from the above work.
The  new  point here is that even the equilibrium field
is large, i.e., the field created  by the  differential rotation which is
itself  strongly reduced by the field in a consistent feedback.
The  
magnetic field decreases fastly near the stellar surface, because differential 
rotation becomes negligible. What is left of the field in the photosphere is difficult
to ascertain without more detailed models. Also, we recall that for consistency
 with Spruit's equations (\cite{Spruit02}),
we have not accounted here for the energy condition  proposed
in paper I (Maeder \& Meynet \cite{Magn1}), which tends to reduce or suppress the field
in the  outer layers. 

\subsection{Diffusion coefficients for the transports of the angular momentum
and  chemical elements}

\begin{figure}[tb]
  \resizebox{\hsize}{!}{\includegraphics{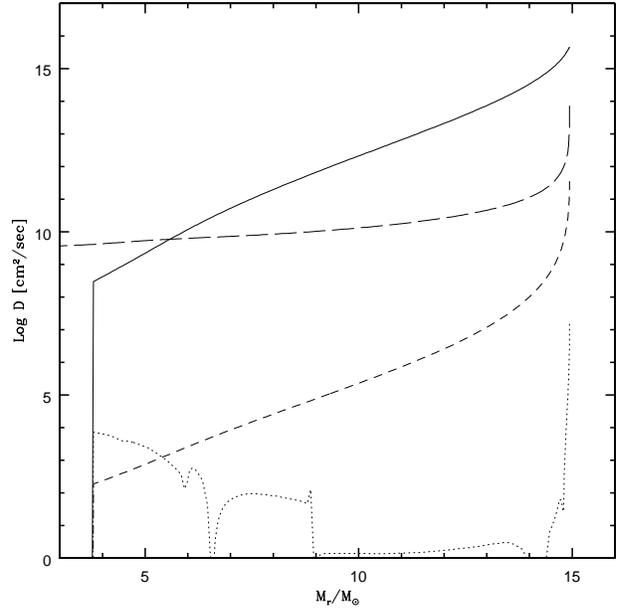}}
  \caption{Internal distribution of the coefficients of diffusion. The coefficient
   $\nu$ (Eq. \ref{nu}) for the transport of angular momentum 
    by the magnetic field is shown by the continuous
  line.                             
   The coefficient   of transport of the chemical elements,
  i.e., the magnetic diffusivity  $\eta$ given by Eq.~(\ref{eta}) is shown by a
  short--dashed line. 
  The coefficient  $D_{\mathrm{shear}}$ of the shear turbulent mixing is shown in comparison
  as a dotted line. The long--dashed line shows the thermal conductivity $K$.
  The ages of the models are the same as 
  in the model of Fig. \ref{NM5}.}  
\label{coeffdiff}
\end{figure}

Fig.~\ref{coeffdiff} shows the coefficients of transport in model M5. The
continuous, higher curves represent the coefficient $\nu$ for 
the transport of the angular momentum 
as given by Eq.~(\ref{nu}). In the outer layers, the value of $\nu$ reaches about
$10^{15}$ cm$^2\cdot$ s$^{-1}$, which differs  by less than one order 
 of magnitude from the typical values  
of the diffusion coefficient in a convective zone. In the region where the
$\mu$--gradient becomes significant, the diffusion is reduced   to  $\nu \approx
10^{9}$ cm$^2\cdot$s$^{-1}$, which is still large.
These  values of $\nu$ implies a strong coupling of the rotation motions,
 which is responsible for the 
near constancy of the angular velocity $\Omega$ through the whole  stellar interior. 
 As illustrated by Eq.~(\ref{nu}) and (\ref{nuzero}), this coefficient
behaves like $(\Omega/N_{\mu})^4$ in this region, which explains the
observed decline towards the interior. This decline is also favoured by the fact that $N^2_{\mathrm{T}}$ also decreases  towards
the deep interior as illustrated by Fig.~\ref{NM5}. We see that $\nu$ and the thermal diffusivity $K$ are of the same order of magnitude in the region above the core, while  $\nu$ is larger
in the outer layers.  This behaviour is consistent with the comparison (cf. Table 1 below)
 of the velocities  of meridional circulation and of the growth of the magnetic instability.

\begin{figure}[tb]
  \resizebox{\hsize}{!}{\includegraphics[angle=0]{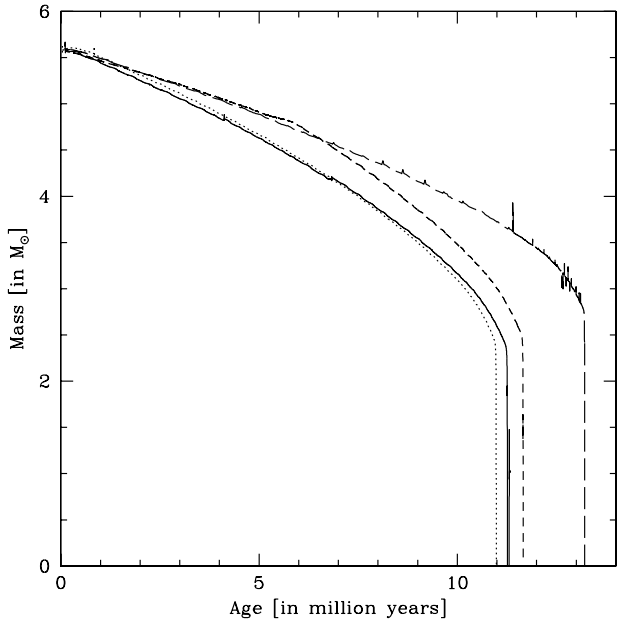}}
  \caption{The
  evolution of the mass of the convective core. The dotted--line
  applies to  model V0, the long--broken (upper) line to model V3, the short--broken
  line to model M1, the continuous line to model M5.}
\label{Mcc}
\end{figure}

The coefficient of magnetic diffusivity  $\eta$, which acts both for
 the diffusion of the magnetic field as well as for the transport of the chemical elements
 as discussed in Sect.~3, is shown by the dashed--line in Fig.~\ref{coeffdiff}.
 Close to the core,   the dependence of
  $\eta_{0}$ (Eq.~\ref{etaz0}) goes with the power of $-6$ on $N_{\mu}$.
  This explains why in the deep interior, near the core,  this coefficient becomes very low, being 
  less than $10^3$  cm$^2\cdot$ s$^{-1}$.
  These regions where $\eta$ is very small tend to inhibit  the transfer of the chemical 
 elements towards the surface
 The coefficient $\eta$ grows rapidly in the very outer layers, rising up 
 to  $10^{12}$ cm$^2\cdot$ s$^{-1}$. We may check here the very small values of the ratio
 $\eta \over K$ mentioned above.

 A curve with coefficient $D_{\mathrm{shear}}$ due to shear turbulence is also shown
 in Fig.~\ref{coeffdiff}.
 As we have seen, the $\Omega$--gradient is not strictly zero. 
 The value of $D_{\mathrm{shear}}$ is negligible
 in the outer layers, while it is larger than  the coefficient for magnetic transport $\eta$
 in the layers close to the core. The reason is that the dependence in 
 $N_{\mu}$ is weaker for $D_{\mathrm{shear}}$ than for $\eta$,
  (being in $N_{\mu}^{-2}$ for the first and in $N_{\mu}^{-6}$ for the second).
   This means that both shear mixing and
 magnetic effects have to be taken into account. It is clear that the interaction of 
 shear instability and magnetic field may be rather complex physically with some
 important feedback effects, particularly in regions where $D_{\mathrm{shear}}$
 and $\eta$ are of the same order. The present work does not solve this point, but 
 Fig.~\ref{coeffdiff} shows the  orders of magnitude  of the main parameters.
 The main conclusion is
 that magnetic coupling dominates the dynamics of rotation. This conclusion is in agreement 
 with that by Heger et al.(\cite{Heger04}), who also find an essentially solid body during MS 
 evolution. We also notice that from a different physical dynamo model
 MacDonald \& Mullan (\cite{MacD04}) suggest that shear instability is present
  over most of the interior, a result which  possibly supports the simultaneous
  consideration of shear transport and magnetic field.

\subsection{Evolution of the core, lifetimes and  tracks in the HR diagram}

Fig.~\ref{Mcc} shows the evolution of the mass of the
convective core. 
As well known, rotation
produces an increase of the core mass fraction and of the MS lifetime. Here, the MS
lifetime of model V3 is about 20 \% larger than for model V0. The models 
M1 and M5 with both
rotation and magnetic field  have a core size and a MS lifetime which are intermediate
between those of models V0 and V3, but closer to the non--rotating model V0.
This is consistent with the results of Figs.~\ref{xE0} and \ref{xM5},
 which show that there is more H  burnt in models
with rotation only. The mixing of the chemical elements by Tayler
instability increases only very slightly the fuel reservoir with respect to a model without
rotation, since the strong inhibition (Eq.~\ref{eta}) by the $\mu$--gradient
limits very much the importance of the mixing. Again, we notice some small
differences between models M1 and M5. The core is larger in M1 than in M5. The
reason is likely due to the slightly larger values of  $D_{\mathrm{shear}}$ 
in model M1 close to the core. The deviation of model M5 from the non--rotating case
is very small: the lifetime is increased by only 2.4 \%.

Fig.~\ref{HR} shows the evolutionary tracks in the HR diagram. In interpreting
these tracks, one must remember that the effects of the
 distortion of the stellar surface by rotation 
have been accounted for in the definition of the $T_{\mathrm{eff}}$. (In models with
rotation, we define  the  average effective
temperature as the stellar luminosity divided by the real distorted
surface of the rotating star, cf. Meynet \& Maeder \cite{MMI}).
This definition is equivalent to seeing
the star with an average  orientation, i.e.,  with an angle $i$ 
between the rotation axis and the line of sight equal to the root
of the second Legendre polynomial, i.e.,  about 54 degrees. This is
why all models with rotation start their evolution  at a place in the HR diagram, 
which is different from that of the model V0  with no rotation.

\begin{figure}[t]
  \resizebox{\hsize}{!}{\includegraphics[angle=-90]{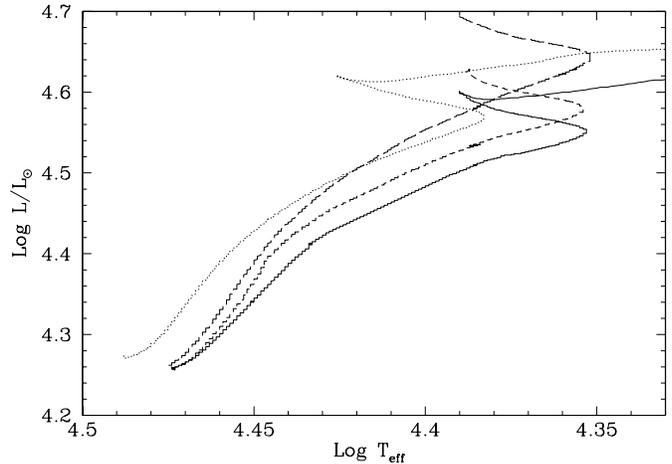}}
  \caption{ Evolutionary tracks in the HR diagram.  The dotted line refers
  to model V0, the long--broken line to model V3, the short--broken
  line to model M1 and the continuous line to model M5. }
\label{HR}
\end{figure}

 Model V3 experiences a larger increase of its luminosity than 
 model V0, as a result of the 
more extended mixing and larger core as shown in Fig.~\ref{xE0}. The growth of 
luminosity for models M1 and M5 is smaller than for model V3,
 because mixing is  less extended (cf. Fig.~\ref{xM5}).
 Models M1 and M5 experience about the same increase in luminosity
 as model V0. The end of the MS  in these two models 
 is a bit more shifted to the red by the atmospheric
 distortion than the zero--age models, since the ratio
 $\Omega \over \Omega_{\mathrm{c}}$ increases from about 0.66  to 0.82 during 
 MS evolution.
 
We see a difference of the tracks for models M1 and M5. A rough agreement 
is achieved with the limiting cases  by Spruit (\cite{Spruit02}), but the
differences are not entierly negligible.
 We also notice the ``stair--case'' track of the magnetic models, particularly 
for model M1. This is the result of the rather particular evolution of the core 
mass of this model as illustrated in Fig.~\ref{Mcc}: initially model M1
 evolves close to the model V3 with rotation, then it behaves more 
 as the model V0 with no rotation, account being given to the
atmospheric effect. This behaviour is rather different for
 model M5 which evolves 
like the model without rotation. This   illustrates the different 
results obtained from  the asymptotic relations and the general solution.

\subsection{Evolution of the abundances of helium and CNO 
elements at the stellar surface}

 \begin{figure}[t]
  \resizebox{\hsize}{!}{\includegraphics[angle=-90]{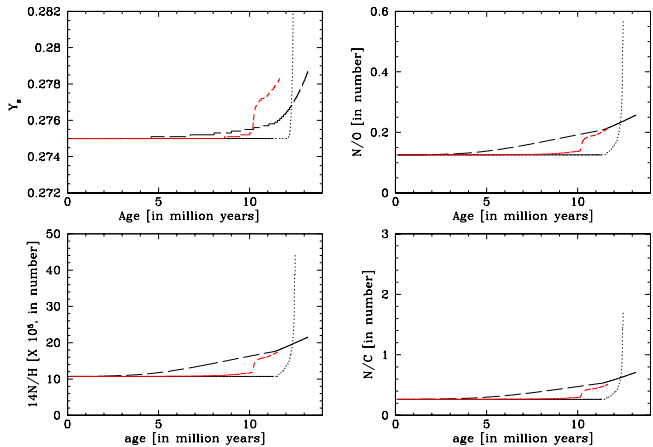}}
  \caption{Time evolution of the surface helium content $Y_{\mathrm{s}}$ in mass fraction, of the 
  N/O, N/H and N/C in mass fraction for the various models V0 (dotted line), 
 V3 (long--broken line), M1 (short--broken line) and M5 (continuous line).    }
 \label{Xi}
\end{figure}

Fig.~{\ref{Xi} reflects the evolution of the abundances of
 helium and CNO elements at the surface of the various models.
 For model V0, as is well known the enrichment only occurs when convective dredge--up appears
 in the red--supergiant stage. For model V3, the growth of the nitrogen
 at the stellar surface amounts to about a factor of 2 during the MS
 phase, as a result of shear mixing mainly and of meridional circulation to a smaller
 extent. The growth of the helium content is quite small and amounts to about
 0.004 in $Y_\mathrm{s}$. 
 Model M1 shows a late N--enrichment up to the level of model V3 near the end
 of the MS, while it is negligible before. The helium enhancement is even higher
 than for model V3.  The striking point is that model M5 
  shows no enrichment at all. This is likely a consequence of slightly lower
  $D_{\mathrm{shear}}$ close to the core than for model M1.
  There the difference between models M1 and M5 is rather critical
  in view of the observational tests to be performed.
  
  We know from spectroscopic observations that significant surface 
  enrichments in helium and nitrogen
  occur in rotating O--stars as well in some early B--type stars (cf. Howarth \&
  Smith \cite{Howarth01}; Villamariz et al. \cite{Villamariz02}). Thus, the fact 
  that the rotating model M5 does not produce such enrichments is a negative
  indication, suggesting that the present models with magnetic field 
  do not well correspond to reality. The fact that the agreement is better
  realized with model M1 is not meaningful, since this results 
  from the use of asymptotic relations
  which lead to an overestimate of the diffusion coefficients in some parts of the star.
  Incidentally, this also shows that wrong conclusions could be drawn,
  if  the general expressions are not used. The models with only rotation much
  better agree with observations (cf. for example Maeder \& Meynet \cite{MMVII}).
   However, one must be careful. These comparisons do not necessarily exclude 
   the presence of any magnetic fields  playing a role in massive stars. In that
   respect, the remarks made  in Sect.~6 below on the equilibrium reached between
   meridional circulation and magnetic transport certainly leads to more 
   differential rotation and mixing than in the present models.

 \section{Discussion: interaction between Spruit's dynamo and meridional circulation.}
 
 The aim of this section is to discuss the  role
 of the magnetic field with respect to other instabilities, in particular the thermal instability, 
  a point which has 
 generally not been  considered. The various authors, for example  Heger et al.
 (\cite{Heger04}), usually just sum up the various effects.
 This simple approach may  nevertheless have the interest to show the order of magnitude of the 
 effects considered. When differences of several orders of magnitude appear between, for example,
 the coefficients for the transport of angular momentum $\nu$ and $D_{\mathrm{shear}}$,
 we may likely reach a  conclusion about which  effect  dominates.
 In this context, we recall that Charbonneau 
 (\cite{Charb01}) has examined the delicate interplay between dynamo and meridional
 circulation. With a different dynamo model, he has concluded that the dynamo action
 remains probable in presence of meridional circulation.
  Our conclusion below is not different. If only a dynamo is present, 
 the star reaches solid body rotation, which would drive a circulation. On the contrary,
 we will see that, if we ignore the dynamo, the meridional circulation and star evolution
 build a differential rotation sufficient to make the dynamo active. Therefore, we may
 conclude that in the  time evolution both effects of  dynamo and meridional 
 circulation are present, with a delicate balance between them. This first 
 approach nevertheless still leaves room for a thorough analysis of the magnetic and thermal
 instabilities at a fundamental level.

 We have seen in paper I (Maeder \& Meynet \cite{Magn1}) that, if in the course of
 evolutionary models calculated with only the current effects of
  rotation (hydrostatic effects, meridional circulation,
   shear mixing, horizontal turbulence, etc...) 
   and which have a significant amount of differential rotation,
  we  estimate the magnetic field created by Tayler--Spruit dynamo,
   then the magnetic field and its transport effects are found  to be very large. 
 The transport coefficients $\nu$ and $\eta$ by the magnetic field were found to 
 overcome the above mentioned effects of rotation, in particular meridional circulation
 which is usually the most important process for the transport of angular momentum.
  This lead us to suggest that magnetic fields may be
 dominating. The situation is probably not so simple.

 In Sect.~5 above, we have followed the evolution of a 15 M$_{\odot}$ star
 with  the magnetic field created
 by  the Tayler--Spruit dynamo
 and we have found an internal equilibrium of rotation 
 characterized by a nearly constant $\Omega(r)$ in the 
 stellar interior due to the strong magnetic coupling expressed by the coefficient $\nu$.
 We may now wonder whether in these models with $\Omega \approx const.$
 the effects of the  magnetic field are always
 larger than those which could be due to the thermal instability  driving
 meridional circulation. If the answer is ``yes'', then clearly Tayler--instability
 and magnetic fields largely dominate  the thermal instability responsible 
 for meridional circulation. If on the contrary the answer is ``no'', this means
 that the situation is more complicated. 
 
 The above worry is quite justified, since we know that there are various feedback
 mechanisms
  playing  a role in the problem. Models with only rotation lead to
 a strong differential rotation, which in turn may feed  the dynamo 
 mechanism and the build--up of a magnetic field, which then would tend to
 dampen the differential rotation, which is its source.
  We also know from previous work
 (Meynet \& Maeder \cite{MMV}) that if rotation is constant in the interior, as is the case 
 in the initial stellar models on the ZAMS, then the thermal instability and
 the resulting velocity $U(r)$ of meridional circulation is much larger 
 than in the case for the equilibrium profile  $\Omega(r)$ of rotation.  
 Moreover $U(r)$ is positive everywhere in the envelope. This means that
 angular momentum is transported from the outer regions of the star to the inner ones, counterbalancing the effects
 of the transport by diffusion. Thus this large
 meridional circulation will enhance  differential rotation. The complete
 feedback loop is the following, with two parts A and B:\\ 
 
 \noindent
 \underline{A) Build--up of magnetic field and constant rotation:}\\
 
 \noindent
 \hspace*{4mm} differential rotation$\quad \Rightarrow \quad$dynamo$\quad\Rightarrow \quad$\\
 \hspace*{4mm} magnetic field$\quad \quad  \;\quad\Rightarrow \; \;  \Omega \sim const.$\\
 
\noindent 
\underline{B) Build--up of  circulation and differential rotation:}\\

\noindent
 \hspace*{4mm}$\Omega \sim const. \quad \; \Rightarrow \quad$stronger thermal instability$\quad \Rightarrow$\\
 \hspace*{4mm}higher $U(r)$$\quad \; \Rightarrow \quad$differential rotation$\quad \Rightarrow$ ...\\

\noindent
The basic equation governing the evolution of $\Omega(r)$ is the equation 
describing the conservation of the angular momentum, which is in  Lagrangian coordinates
(cf. Zahn, \cite{Zahn92}; Maeder \& Zahn \cite {MZ98})

\begin{eqnarray}
\frac{d}{d t} \left(\rho r^2 \overline{\Omega}\right)_{M_r} =
\frac{1}{5 r^2}  \frac{\partial}{\partial r} \left(\rho r^4 \overline{\Omega}
  U(r)  \right)\nonumber \\[2mm]
 \qquad + \frac{1}{r^2} \frac{\partial}{\partial r}
\left(\rho \; \nu \; r^4 \frac{\partial\overline{\Omega}}{\partial r} \right) \; .
\label{angmomentum}
\end{eqnarray}

\noindent
$\overline{\Omega}(r)$ is the average angular velocity on the equipotential.
If an equilibrium  of the above feedback loop is reached in a time relatively 
short with respect to the evolutionary timescale, this equation is equal to zero.
 We would thus have a balance of the flux due to
meridional circulation $U(r)$ by the flux due to the magnetic diffusion expressed by $\nu$,
as shown by Eq.~(\ref{angmomentum}).The equilibrium is achieved for 

\begin{equation}
U(r) = 
\; 5 \;  \frac{\nu}{r} \; \left(- \frac{\partial \ln \Omega}{\partial \ln r} \right) \; .
\label{U5}
\end{equation}

\noindent
The factor 5 results from the integration of the angular momentum over 
the isobar,  which leads to  the above Eq.~(\ref{angmomentum}).
On the right  side of  Eq.~(\ref{U5}), we have a velocity which can be
associated with the magnetic diffusion. Let us call it $U_{\mathrm{magn}}$ and
the left member $U_{\mathrm{circ}}$ to make the distinction. In Table 1, we compare 
the values of $U_{\mathrm{magn}}$ and $U_{\mathrm{circ}}$ at the middle of the MS phase
in the model of 15 M$_{\odot}$ studied in Sect.~5.

  \begin{table}[htbp]
\caption{Comparison between $U_{\rm circ}$ and $U_{\rm magn}$ in
 a 15 M$_{\odot}$ model at X$_{\rm{c}}=0.30$.} \label{tb1}
\begin{center}\scriptsize
\begin{tabular}{cccccc}
$M_r/M_\odot$   & $U_{\rm circ}$  & $\nu$ & $|q|$ & $\log r$ & $U_{\rm magn}$   \\
    &                   &          &           &           &               \\
\hline
    &                   &          &           &           &               \\
 8.02 & 1.97e-02  & 2.03e+11  & 0.45e-3  & 11.1060 & 3.60e-03 \\
10.05 & 1.55e-02  & 2.24e+12  & 0.62e-4  & 11.1897 & 4.46e-03 \\
13.04 & 1.85e-02  & 7.78e+13  & 0.49e-4  & 11.3298 & 9.00e-02 \\

    &                   &          &           &           &               \\        
\hline                  
\end{tabular}
\end{center}
\end{table}

\noindent
We see that $U_{\mathrm{circ}}$ dominates over $U_{\mathrm{magn}}$ in the deep
envelope, while the opposite situation occurs
closer to the surface;
the difference reaches about a factor of 5. 
Models with only rotation  as in paper I indicate that the magnetic
fields should be large, while models with magnetic fields as here
indicate that the meridional circulation is significant. The basic reason for the
difference is the $\Omega(r)$ profile, with important differential rotation in the first case
and constant rotation in the second.

We  think that an equilibrium profile of $\Omega(r)$
is reached as a result of the interaction of the two effects,
 neither  with  as much differential rotation as in the models 
with rotation only (cf. Fig.~1), nor with a constant angular velocity
as in the present models with magnetic field (cf. Figs.~2 and 3). This
intermediate $\Omega(r)$  profile will lead to more mixing than in the present case.
Will this be enough to be in agreement with the observations? 
We do not know yet. We emphasize that the equilibrium
profile of $\Omega(r)$ is likely defined by a particular balance
between $U_{\mathrm{magn}}$ and $U_{\mathrm{circ}}$, with account taken also
of the stellar contraction and angular momentum conservation.
There is probably an additional constraint imposed by the fact that the
energy of the magnetic field originates from the excess energy in differential 
rotation and thus the energy density in the magnetic field is limited by its 
energy source. Such effects will be examined in further work.


 \section{Conclusions}

 The equations for the general case of the Tayler--Spruit dynamo 
 with $\mu$--gradients and nonadiabatic effects 
 have been developed. The general expressions  consistently
 allow us to recover Spruit's asymptotic relations (\cite{Spruit02}), when  $N_{\mu}$ or $N_{\mathrm{T}}$ 
 are zero.  Numerically the general and asymptotic results are consistent
  with some   quantitative differences. This demonstrates 
   the need to use the general
  solutions, in particular 
  if we want to perform observational tests. 
 
  Some initial results on the role  of possible magnetic field in stellar models
  have been found. The Tayler--Spruit dynamo 
  imposes that the stars to rotate nearly as a solid body. This leads to a different
  evolution of rotational velocities as stars move away from the zero--age
  sequence. In general, the rotation velocities are  higher  
 when a magnetic field is present. A second general result is that internal mixing
 and surface enrichments in products of the CNO burning  are 
 smaller when a magnetic field is present than in rotating models without a 
 magnetic field. Since spectroscopic observations of OB stars
 show significant N-- and He--enrichments, this  cast some doubt as to
 whether magnetic fields have a dominant effect, especially more than the models 
 which have only rotation and no magnetic field show a good agreement with observations. 
  
  If magnetic fields are present, we may expect, as shown in Sect.~5, that 
  the magnetic instability and the thermal instability which drives meridional circulation
  reach some balance in their complex feedback loop. For reasons explained in Sect.~5, this
  balance should be characterized by less differential rotation 
  than in current models with only rotation  and by more
 differential rotation than in the present  models with a magnetic field.
 Such models may lead to more mixing than the present ones.

  Finally, we emphasize that  at this stage
 we really do not know how the Tayler--Spruit instability interacts
 with the meridional circulation and how  the energy constraints
 may modify the results.
 But we have explored some consequences 
  of the magnetic field in the absence of  meridional circulation and examined
  the possibilities  for its existence.



\end{document}